\documentstyle[preprint,aps]{revtex}

\begin{document}
\draft

\title{On finite--temperature and --density radiative corrections
to the neutrino
effective potential in the early Universe}

\author{R.Horvat \\
   ``Rugjer Boskovic'' Institute, P.O.Box 1016, 10001 Zagreb,
Croatia.}

\maketitle

\begin{abstract}
Finite--temperature and --density radiative corrections to the neutrino
effective potential in the otherwise $CP$--symmetric early Universe are
considered in the real--time approach of Thermal Field Theory. A consistent
perturbation theory endowed with the hard thermal loop resummation
techniques developed by Braaten and Pisarski is applied. Special attention is
focused on the question whether such corrections can generate any nonzero
contribution to the $CP$--symmetric part of the 
 neutrino potential, if the contact approximation for the
$W$--propagator is used.
\end{abstract}


    For neutrinos propagating in matter, the vacuum energy--momentum
relation is no longer respected.The modification of the neutrino dispersion
relation can be represented in terms of an index of refraction \cite{1} or an
effective potential \cite{2}. However, in the framework of Thermal Field
Theory (TFT) it is based on the real part of the matter--induced neutrino
self--energy \cite{3}.

    The subject of neutrino propagation in matter became popular firstly
when Wolfenstein calculated neutrino refractive index in matter
\cite{4}. Later on, Mikheyev and Smirnov recognized resonant nature of flavor
oscillations triggered by matter effects \cite{5}. The hypothesis  based on
these effects has even become the most popular explanation for the
solar--neutrino deficit \cite{6}.

    Neutrino oscillations could also be cosmologically important. The
oscillations between a standard left--handed neutrino $(\nu_e,\, \nu_{\mu}\,
or\,
\nu_{\tau})$ and a $SU(2)_L$
singlet $\nu_S$ has attracted considerable attention recently
\cite{7}. Oscillations into new neutrino states would distort successful Big
Bang Nucleosynthesis (BBN) and hence a bound on oscillations parameters can
be derived. The constraints on mixing between active and sterile neutrinos
are such that excluded regions include the large angle  $\nu_e-\nu_S$ MSW
solution
\cite{8} as well as the $\nu_{\mu}-\nu_S$ mixing solution \cite{9}
to the atmospheric
neutrino problem, provided that there was no significant $CP$ asymmetry
in the early
Universe. In addition, for Dirac neutrinos endowed with anomalous magnetic
dipole moments, the magnetically induced neutrino oscillations between
left--handed $(\nu_L)$ and right--handed $(\nu_R)$ states may occur \cite{10}.
In such
oscillations the BBN arguments constrain the product of neutrino MDM's and a
present--day intergalactic field strength.  

     In all above situations in the early Universe (and also in other
environments) the neutrino refractive effects are of crucial importance. In
the zero--temperature and --density (0TD) limit, it suffices  to calculate
an index of refraction, obtained directly from the neutrino
forward--scattering amplitude. For lowest order calculations at
finite--temperature and --density (FTD), one usually uses the old version
of TFT formulated in the real time, as developed first by Dolan and Jackiw
\cite{11} (in the real time formalism it is more convenient to separate the
vacuum effects from the effects of the medium). However, this old version is
not applicable to higher--order calculations at FTD, since it is plagued
with ill--defined pinch singularities which do not cancel. Instead, one
should use a consistent TFT perturbative approach initiated by Niemi and
Semenoff \cite{12}, and amplified later on by Braaten and Pisarski \cite{13}
through the resummation program for the soft regime.  

     In terms of standard electroweak interactions the almost $CP$ symmetry
of the medium implies that the lowest order refractive effects of order
$G_FT^3$,
which dominate in stars, nearly cancel. However, the $CP$--symmetric
contribution, which arises from an expansion of gauge--boson propagators,
turns out to be suppressed by a very small factor of order $T^2/M^2_W$,
where the
temperature $T$ in the epoch of interest $T\,\sim\,0.1-10$ MeV.
Since the only nonzero
contribution is of order $G_FT^5$, the smallness of such tree--level
result calls
for an investigation of its radiative corrections. It was found \cite{14}
that the 0TD radiative correction to the neutrino index of refraction in the
early Universe is about $20\%$ for $\nu_e$ and $50\%$ for $\nu_{\mu}$ and
$\nu_{\tau}$. The FTD radiative
corrections have not been considered yet, except the general proof for
cancellation of infrared and mass (or collinear) singularities at order
$\alpha$, as given recently in \cite{15}. In the present note,
we shall investigate
FTD radiative corrections to the lowest order result (the zeroth
order in the expansion of the $W$--propagator, i.e., the contact approximation),
to see if they could produce any term not proportional to the
particle--antiparticle asymmetry, namely of order $\alpha G_FT^3$.
The main difference
between the 0TD and FTD radiative corrections is the appearance of thermal 
$(e^{\mp},\gamma)$
loops in the latter, characterized by circulation of real particles from the
heat bath. Thus, for example, the FTD radiative correction to the process
$\nu_ee^{\mp}\longrightarrow\nu_ee^{\mp}$
brings in some extra processes like
$\nu_ee^{\mp}\gamma\longrightarrow\nu_ee^{\mp}\gamma$,
$\nu_e\gamma\longrightarrow\nu_e\gamma$,
 where $\gamma\,'s$  are from the heat bath.

       To begin, let us remind how the cancellation in the lowest order
contribution results. We will be always working  in the limit of perfect $CP$
symmetry, which, as a consequence, results in the equal amounts of particles
and antiparticles in the early Universe (stated differently, the chemical
potential $\mu=0$ is assumed for all the species). Then, the standard bubble
charged--current graph for $\nu_e$ gives the following contribution to the
neutrino self--energy in the contact approximation, 
\begin{equation}
\Sigma^{(W)}_{st.}=2\sqrt{2}\, G_F
\int \frac{d^4k}{(2\pi)^4}
\gamma^\mu L i S_{11}(k) \gamma_\mu L~,
\end{equation}
 where the (11) component of the real--time electron propagator is given by
\begin{equation}
iS_{11}(k)=(\not{\!k} + m_e)\,
\left[\frac{i}{k^2 - m_e^2 +i\varepsilon} - 2\pi\delta(k^2 - m_e^2)
\sin^2\!{\phi}_{k}\right]~,
\label{S11}
\end{equation}

\noindent and 

\begin{equation}
\sin^{2}\!\phi_{k}=\frac{1}{e^{\beta|k_0|} + 1}
\end{equation}             
is the Fermi--Dirac distribution function. Insertion of the FTD part of (2)
into (1) gives
\begin{equation}
\Sigma^{(W)}_{st.}=4\sqrt{2}\, G_F
\left \{ \int \frac{d^3k}{(2\pi)^3}
\int_{-\infty}^{+\infty}dk_0k_0\delta(k_0^2 - k^2 - m_e^2)
\frac{1}{e^{\beta|k_0|} + 1} \right \}
\gamma_0 L~~,
\end{equation}
where $k\equiv|\vec{k}|$.
In Eqs.(3) and (4), the four--velocity of the center of mass of the medium
$u^{\mu}$ is given by its value in the rest frame of the medium;
$u^{\mu}=(1,\vec{0})$.
The effective potential is obtained from a neutrino self--energy by
omitting the Dirac part $\not{\!u}L$. In a mathematical sense, it is easy
to understand
why Eq.(4) vanishes identically:We simple integrate the odd function of
$k_0$
over the symmetric interval. Now, we are going to check up whether a
mathematical feature like that survives FTD radiative corrections.

        The correction to the first order (in $G_F$) FTD contribution to the
self--mass of $\nu_e$ due to photon radiation amounts to calculate the $\cal
O(\alpha)$
correction to the bare electron propagator. This is performed using the
Schwinger--Dyson equation for the full fermion propagator. In our analysis,
for simplicity, the electron is always considered to be massless, i.e.,
$m_e=0$, a
standard assumption in the early Universe calculations around the BBN
epoch. Summing over different internal vertices and using the
``momentum--derivate formula'', the final result in the real--time framework
can be written in the following form 
\begin{eqnarray}
i\delta S_{11}^M(k) &=& \lim_{{\hat{k}}^2 \rightarrow k^2}
\left \{ \pi \frac{\partial}{\partial {\hat{k}}^2} \delta
(k_0^2-{\hat{k}}^2) \not{\!k} \mbox{\rm Re}\Sigma_M(k_0,\vec{k})\!\!\!\not{\!k} -
PP\frac{1}{(k_0^2-{\hat{k}}^2)^2}\not{\!k}
\mbox{\rm Im}\Sigma_M(k_0,\vec{k})\!\!\!\not{\!k}
\right \}  \nonumber  \\
&-& 
2\lim_{{\hat{k}}^2 \rightarrow k^2}\left \{ \pi \frac{\partial}{\partial
{\hat{k}}^2} \delta(k_0^2-{\hat{k}}^2) \not{\!k}\!
\left [\mbox{\rm Re}\Sigma_0(k_0,\vec{k})+\mbox{\rm Re}
\Sigma_M(k_0,\vec{k})\right ]\!\! \not{\!k} 
\right. \nonumber  \\
&-& \left.  PP\frac{1}{(k_0^2-{\hat{k}}^2)^2} \not{\!k}\! \left[
\mbox{\rm Im}\Sigma_0(k_0,\vec{k})+\mbox{\rm Im}\Sigma_M(k_0,\vec{k})\right]
\!\!\not{\!k} \right \}
 \sin^2\!\phi_k
\nonumber \\
&+&
\lim_{{\hat{k}}^2 \rightarrow k^2} \left \{ i\pi \frac{\partial}{\partial
{\hat{k}}^2} \delta
(k_0^2-{\hat{k}}^2) \not{\!k} \mbox{\rm Im}\Sigma_M(k_0,\vec{k})\!\!\not{\!k} -
PP\frac{i}{(k_0^2-{\hat{k}}^2)^2}\not{\!k}
\mbox{\rm Re}\Sigma_M(k_0,\vec{k})\!\!\not{\!k}
\right \}~.  
\end{eqnarray}       
Since it is assumed that $m_e=0$, the well known ``mass--derivative formula''
\cite{16} has to be replaced with the ``momentum--derivative formula''. Since
we want to
 discuss the effects of the medium, only FTD parts are kept in
(5). Also, note that the terms in the third curly bracket in (5) do not
contribute to the real part of the neutrino self--energy. Moreover,
we have split in
(5) the electron self--energy function $\Sigma$ into the vacuum and the matter
part. The parts associated to Re$\Sigma_0$, Re$\Sigma_M$ contain virtual photon corrections (at
0TD and FTD, respectively) and correspond to a cut through the internal
line. On the other hand, the parts involving Im$\Sigma_0$,
Im$\Sigma_M$ correspond to a cut
through $\Sigma$ and are associated to a real process. Finally,
Eq.(5) is understood
as integrating  over the $\delta$--function before taking the derivative.

     If $m_e=0$, the electron mass operator can be decomposed as
\begin{equation}
\Sigma(k_0,\vec{k}) = a(k_0,k)\gamma^0 + b(k_0,k)\vec{k}\cdot\vec{\gamma}~~.
\end{equation}
Then, with the aid of generalized Cutosky rules at FTD \cite{17}, we obtain
for the imaginary parts,

\begin{mathletters}
\begin{eqnarray}
\mbox{\rm Im}[a_0(k_0,k) + a_M(k_0,k)] &=&
(4\pi\alpha\varepsilon(k_0)/\sin{2\phi_k})
\nonumber \\
&\times&
\left \{ 
   \int_{\frac{1}{2}(k_0-k)}^{\frac{1}{2}(k_0+k)}
   \frac{xdx\Theta(x)}{32\pi k}\,
   \frac{\varepsilon(k_0-x)}
          {\sinh \left[ \frac{1}{2}\beta(k_0-x) \right] 
           \cosh(\frac{1}{2}\beta x)}
\right.
\nonumber \\
&+&  
\left.
   \int_{-\frac{1}{2}(k_0+k)}^{-\frac{1}{2}(k_0-k)}    
   \frac{xdx\Theta(x)}{32\pi k}\,
   \frac{\varepsilon(k_0+x)}
          {\sinh \left[ \frac{1}{2}\beta(k_0+x) \right]
           \cosh(\frac{1}{2}\beta x)} 
\right \}~, 
\end{eqnarray}

\begin{eqnarray}
\mbox{\rm Im}[b_0(k_0,k) + b_M(k_0,k)] &=&
(4\pi\alpha\varepsilon(k_0)/\sin{2\phi_k})
\nonumber \\
&\times&
\left \{ -\frac{k_0}{k^2}\left[
   \int_{\frac{1}{2}(k_0-k)}^{\frac{1}{2}(k_0+k)}
   \frac{xdx\Theta(x)}{32\pi k}\,
   \frac{\varepsilon(k_0-x)}
          {\sinh \left[ \frac{1}{2}\beta(k_0-x) \right]
           \cosh(\frac{1}{2}\beta x)} \right. \right. \nonumber \\
&+&  \left. \int_{-\frac{1}{2}(k_0+k)}^{-\frac{1}{2}(k_0-k)}
   \frac{xdx\Theta(x)}{32\pi k}\,
   \frac{\varepsilon(k_0+x)}
          {\sinh \left[ \frac{1}{2}\beta(k_0+x) \right]
           \cosh(\frac{1}{2}\beta x)} \right]
\nonumber \\ 
&+& \left(k_0^2/2k^2-1/2 \right) \left[
   \int_{\frac{1}{2}(k_0-k)}^{\frac{1}{2}(k_0+k)}
   \frac{dx\Theta(x)}{32\pi k}\,
   \frac{\varepsilon(k_0-x)}
          {\sinh \left[ \frac{1}{2}\beta(k_0-x) \right]
           \cosh(\frac{1}{2}\beta x)} \right.
\nonumber \\
&-&  \left. \left. \int_{-\frac{1}{2}(k_0+k)}^{-\frac{1}{2}(k_0-k)}
   \frac{xdx\Theta(x)}{32\pi k}\,
   \frac{\varepsilon(k_0+x)}
          {\sinh \left[ \frac{1}{2}\beta(k_0+x) \right]
           \cosh(\frac{1}{2}\beta x)} \right]
\right \}~.
\end{eqnarray}
\end{mathletters}

\noindent Also, trivially, we have 

\begin{mathletters}
\begin{equation}
\mbox{\rm Im}a_0(k_0,k) = -\frac{\alpha}{4}k_0\Theta(k_0^2-k^2) ~,
\end{equation}
\begin{equation}
\mbox{\rm Im}b_0(k_0,k) = \frac{\alpha}{4}\Theta(k_0^2-k^2)~.
\end{equation}
\end{mathletters}

 The easiest way to account for the terms involving
Re$\Sigma_M$\footnote{Within the above
formalism, it can be easily seen that the term involving the renormalized
part of Re$\Sigma_0$ vanishes, in agreement with the results obtained earlier
\cite{17}.} is by
writing down a dispersion relation for the matter part of the self--energy
at both poles,
\begin{equation}
\mbox{\rm Re}a_M(\pm k_0,k) = \pm \frac{PP}{\pi} \int_{-\infty}^{+\infty}
dk_0' \frac{\mbox{\rm Im}a_M(k_0',k)}{k_0'\, \mp k_0}~~,
\end{equation}
and the same for Re$b_M(\pm k_0,k)$.
Eq.(9) is based on the fact that $\Sigma_M$ $\sim$ exp$(-\beta|p_0|)$ as
$|p_0|$ $\rightarrow$ $\infty$. Gathering all together, one finds, after
performing angular integrations, 
that the FTD radiative correction at $\cal O(\alpha)$
to the standard result (\ref{S11}) is
obtained by the replacement:

\begin{eqnarray}
& &\int_{-\infty}^{+\infty}dk_0k_0\delta (k_0^2-k^2)\frac{1}{e^{\beta
|k_0|}+1}\rightarrow
\lim_{{\hat{k}}^2 \rightarrow k^2} \left \{ \int_{-\infty}^{\infty}dk_0
\left [ \frac{\partial}{\partial {\hat{k}}^2} \delta(k_0^2-{\hat{k}}^2)
\left [ (k_0^2+k^2)\mbox{\rm Re}a_M(k_0,k)  \right.  \right.  \right.
\nonumber \\
&+& \left. \left. 2k_0k^2\mbox{\rm Re}b_M(k_0,k) \right]
   \left( -\frac{1}{2}+\frac{1}{e^{\beta|k_0|}+1} \right)
\right]  \nonumber  \\
&+& \frac{1}{2\pi}PP\frac{1}{(k_0^2-{\hat{k}}^2)^2}
\left[ (k_0^2+k^2)\mbox{\rm Im}a_M(k_0,k)+2k_0k^2
  \mbox{\rm Im}b_M(k_0,k) \right]
\nonumber \\
&-&
\frac{1}{\pi}PP\frac{1}{(k_0^2-{\hat{k}}^2)^2}  
\left[ (k_0^2+k^2)\mbox{\rm Im}[a_0(k_0,k)+a_M(k_0,k)]  \right.
\nonumber \\
&+& \left. \left. 2k_0k^2\mbox{\rm Im}[b_0(k_0,k)+b_M(k_0,k)] \right]    
\frac{1}{e^{\beta|k_0|}+1}  \right \}~~.
\end{eqnarray}

\noindent By a direct inspection of the above expressions, one concludes that

\begin{mathletters}
\begin{equation}
\mbox{\rm Im}[a_0(-k_0,k)+a_M(-k_0,k)]=-\mbox{\rm Im}[a_0(k_0,k)+a_M(k_0,k)]~~,
\end{equation}
\begin{equation}
\mbox{\rm Im}[b_0(-k_0,k)+b_M(-k_0,k)]=-\mbox{\rm Im}[b_0(k_0,k)+b_M(k_0,k)]~~,
\end{equation}
\begin{equation}
\mbox{\rm Re}a_M(-k_0,k)=-\mbox{\rm Re}a_M(k_0,k)~~,
\end{equation}
\begin{equation} 
\mbox{\rm Re}b_M(-k_0,k)=-\mbox{\rm Re}b_M(k_0,k)~~.
\end{equation}
\end{mathletters}

Hence, upon inclusion of the FTD radiative corrections at $\cal O(\alpha)$, one finds that
one odd function of $k_0$ is replaced by another odd function of  $k_0$, and the
net result is again zero.

      However, strictly speaking, the above analysis is adequate only for
hard loop momenta of order $T(k_0,k \sim T)$, 
whereas for soft momenta $(k_0,k \sim eT)$ the resummation
program developed by Pisarski \cite{19}, Braaten and Pisarski \cite{13} and
Frenkel and Taylor \cite {20} should be applied. The starting point is that,
in the sense of ``hard thermal loop'' \cite{13,20}, one has to make a
distinction between hard loop momenta of order $T$ and soft momenta of order
$eT$. The thermal mass of the electron, being of order $eT$, is generated by a
loop integral where the momentum running inside the loop is hard. The hard
momentum contribution to the thermal self--energy of the electron is called
a hard thermal loop \cite{13,20}. Finally, only soft lines need to be
ressumed (the HTL resummed propagators are used), whereas for hard lines the
bare perturbation series can still be used. Usually, the arbitrary
intermediate energy--momentum cut--off $k_c$ of order $\sqrt{eT}$
is put by hand to
separate the two regimes (the final result should be independent of $k_c$
\cite{21}).

     The matter part of the resummed electron propagator is given by (only
the real part of it is kept)
\begin{equation}
iS_{11}^{M,R}(k_0,\vec{k}) =(1-2\sin^2\!\phi_k)\mbox{\rm Re}\left
(\frac{i}{\not{\!k}-\Sigma(k_0,\vec{k})+i\varepsilon} \right)~~.
\end{equation}
With the explicit expressions for the fermionic HTL (first determined by
Klimov \cite{22}  and Weldon \cite{23}), it takes the form 
\begin{equation}
iS_{11}^{M,R}=\left(-\frac{1}{2}+\sin^2\!\phi_k \right)
\left \{(\gamma_0+\vec{\gamma}\cdot\hat{k})\mbox{\rm Im}\frac{1}{A_0+A_S} +
(\gamma_0-\vec{\gamma}\cdot\hat{k})\mbox{\rm Im}\frac{1}{A_0-A_S} \right \}~~,
\end{equation}
where the two functions with opposite chirality/helicity ratio are given by 
\begin{equation}
\frac{1}{A_0\mp A_S}=\frac{1}{ k_0\mp k-\frac{\textstyle m_e^2}
  {\textstyle 2k}\left[
\left(1\mp\frac{\textstyle k_0}{\textstyle k} \right)
\ln{\frac{\textstyle k_0+\textstyle k}{\textstyle k_0-\textstyle k}} \pm 2
 \right]}~~,
\end{equation}
and $m_e^2=e^2T^2/8$ is the thermal mass of the electron.
Note, that there are four poles in $S_{11}^{M,R}$ \cite{24}. Hence, for soft
lines, the replacement (10) is actually given by

\begin{equation}
\int_{-k_c}^{k_c}dk_0k_0\delta (k_0^2-k^2)\frac{1}{e^{\beta |k_0|}+1}
\rightarrow \int_{-k_c}^{k_c}\frac{dk_0}{2\pi}
\left \{ \mbox{\rm Im} \frac{1}{A_0+A_S}+\mbox{\rm Im}
\frac{1}{A_0+A_S} \right \}
\left( \frac{1}{2}-\frac{1}{e^{\beta |k_0|}+1} \right)~.
\end{equation}

      Above the light cone, the renormalized propagator is determined by
quasiparticles which are collective excitations. We find for $k_0>k(k_0>0)$,
\begin{eqnarray}
\mbox{\rm Im}\frac{1}{A_0+A_S+i\varepsilon} +
\mbox{\rm Im}\frac{1}{A_0-A_S+i\varepsilon}  &=&
(-\pi/2m_e^2)[\delta(k_0-\omega_-(k))(\omega_-^2(k)-k^2)  
\nonumber \\
 &+&
\delta(k_0-\omega_+(k))(\omega_+^2(k)-k^2)]~~,
\end{eqnarray}
and for $|k_0|>k(k_0<0)$ 
\begin{eqnarray}
\mbox{\rm Im}\frac{1}{A_0+A_S-i\varepsilon} +
\mbox{\rm Im}\frac{1}{A_0-A_S-i\varepsilon}  &=&
(\pi/2m_e^2)[\delta(k_0+\omega_+(k))(\omega_+^2(k)-k^2)
\nonumber \\
 &+&
\delta(k_0+\omega_-(k))(\omega_-^2(k)-k^2)]~~,  
\end{eqnarray}
where $\omega_{\pm}$ denote the two dispersion laws \cite{24}.

      Instead, bellow the light cone, the renormalized propagator is
determined by the imaginary part of the HTL, which is nonzero owing to the
Landau damping mechanism \cite{13,19}. Thus, we find for $|k_0|<k$, 
\begin{mathletters}
\begin{equation}
\mbox{\rm Im}\frac{1}{A_0-A_S+i\varepsilon}=
-\frac{\pi(m_e^2/2k)\left(1-\frac{\textstyle k_0}{\textstyle k}
\right)\varepsilon(k_0)}
{{\left(k_0-k-\frac{\textstyle m_e^2}{\textstyle 2k}
\left[ \left(1-\frac{\textstyle k_0}{\textstyle k}
\right)\ln\left| \frac{\textstyle k_0+\textstyle k}{\textstyle
k_0-\textstyle k}
\right|+2\right] \right)}^2
+\frac{\textstyle m_e^2}{\textstyle 4k^2}
\left(1-\frac{\textstyle k_0}{\textstyle k}\right)^2{\pi}^2}~~,
\end{equation}
\begin{equation}
\mbox{\rm Im}\frac{1}{A_0+A_S+i\varepsilon}=-\frac{\pi(m_e^2/2k)
\left(1+\frac{\textstyle k_0}{\textstyle k}  
\right)\varepsilon(k_0)}{{\left(k_0+k-\frac{\textstyle m_e^2}{\textstyle 2k}
\left[
\left(1+\frac{\textstyle k_0}{\textstyle k}
\right)\ln\left| \frac{\textstyle k_0+\textstyle k}
{\textstyle k_0-\textstyle k}
\right|-2\right] \right)}^2
+\frac{\textstyle m_e^2}{\textstyle4k^2}
\left(1+\frac{\textstyle k_0}{\textstyle k}\right)^2{\pi}^2}~~.
\end{equation}     
\end{mathletters}                                      
By noting that
\begin{equation}
\mbox{\rm Im}\frac{1}{A_0\mp A_S+i\varepsilon}(-k_0,k)=
-\mbox{\rm Im}\frac{1}{A_0\pm
A_S+i\varepsilon}(k_0,k)~~,
\end{equation}
we can see that the new function is again odd with respect to $k_0$, and
therefore its contribution vanishes.

      In conclusion, we have considered previously ignored higher--order
corrections to the neutrino effective potential in the early Universe,namely
the FTD radiative corrections at $\cal O(\alpha)$,
in a theory where only the contact part
of the $W$--boson propagator is kept. Searching for $\alpha G_FT^3$ corrections,
we have
applied a consistent TFT in the real time. The FTD radiative corrections
comprise, beside the usual virtual corrections at 0TD, the virtual
corrections at FTD as well as the real corrections. We have found that they
share the same feature as the lowest order result:They vanish in a
CP--symmetric plasma. Because of the resummed character of a perturbation
theory, it is easy to show that the same feature persists to all--order
perturbation contributions. \newline

{\bf Acknowledgements. } The author acknowledges the support of the Croatian
Ministry of Science and
Technology under the contract 1 -- 03 -- 068.

\end{document}